# Investigation of magneto-optical properties of ferrofluids by laser light scattering techniques

E K Nepomnyashchaya[a], A V Prokofiev[a], E N Velichko[a], I V Pleshakov[a,b], Yu I Kuzmin[a,b]

[a]Department of Quantum Electronics, Peter the Great Saint Petersburg Polytechnic University, Saint Petersburg 195251, Russia

[b]Laboratory of Quantum Electronics, Ioffe Institute, Saint-Petersburg, 194021, Russia

elina.nep@gmail.com

**Abstract**. Investigation of magnetooptical characteristics of ferrofluids is an important task aimed at the development of novel optoelectronic systems. This article reports on the results obtained in the experimental studies of the factors that affect the intensity and spatial distribution of the laser radiation scattered by magnetic particles and their agglomerates in a magnetic field. Laser correlation spectroscopy and direct measurements of laser radiation scattering for studies of the interactions and magnetooptical properties of magnetic particles in solutions were employed. The objects were samples of nanodispersed magnetite ($Fe_3O_4$) suspended in kerosene and in water. Our studies revealed some new behavior of magnetic particles in external magnetic and light fields, which make ferrofluids promising candidates for optical devices.

## 1. Introduction

Ferrofluids (FFs) attract considerable attention from the point of view of fundamental research and also possible applications, among which optical devices have recently been discussed [1, 2]. Though ferrofluids were first produced a few decades ago, a lot of questions on their behavior remain to be unanswered. In particular, FFs have nonlinear optical properties which are poorly understood, but which may be used in modern photonic devices [3, 4].

To investigate optical and magnetooptical properties of FFs, the methods based on laser radiation scattering are appropriate. When laser radiation is transmitted through a colloidal medium with magnetic nanoparticles, some important effects can be observed. For example, under the influence of incident light magnetic particles may conglomerate [5]. In addition, external magnetic field can change optical properties of ferrofluids, for example, it can lead to the formation of elongated clusters or chains [1, 6].

In order to understand the behavior of nanoparticles in a ferromagnetic fluid, a number of different physical techniques are invoked. Among them are optical techniques which give valuable information on ferrofluids properties. This paper reports on the studies involving i) measurements of scattered light and a subsequent analysis of its two-dimensional Fourier transform and ii) laser spectroscopy based on the correlation analysis of scattered radiation.

## 2. Materials and Methods

### 2.1. Samples

The objects of our studies were samples of magnetite $Fe_3O_4$ suspended in water and kerosene solvents. The oleic acid was used as a surfactant that prevented particle aggregation. The average size of the particles was about 7–8 nm and typical concentrations were 0.02 and 0.2 vol. %.

The samples of ferrofluids were produced by the method described in [7]. It was shown in it that FF nanoparticles follow a log-normal distribution with a scaling parameter of 0.17. Before measurements the ferrofluids samples were thoroughly mixed, however, as the particle size distributions showed that small amounts of aggregates about 13 nm in size were present in the samples before the field application.

### 2.2. Experimental methods

To obtain 2D pictures of scattered radiation, direct measurements of laser radiation scattering were used (Fig. 1,a). The magnetic field was generated by two magnets and measured by a Hall probe.

The field distribution between the magnet poles was measured by a Hall sensor. It was found that the field gradient at the sample location (in a volume with dimensions of about 5 × 5 × 5 mm at the magnetic system center) was less than 1 Oe/cm. Within the beam the optical field was nearly homogeneous. According to our estimates, such a weak heterogeneity did not exert an appreciable influence on the formation of clusters in ferrofluids.

Laser correlation spectroscopy is an efficient tool for measuring sizes and investigation of cluster formation in solutions [8, 9]. The technique relies on measurements of the autocorrelation function of the light scatted by a sample. The correlation analysis of the signal provides information on translational and rotary Brownian diffusion. To obtain the correlation curve, the experimental setup shown in Fig. 1, b was built. In this setup, a coherent linearly polarized light beam from a laser with a diameter of 1.2 mm was transmitted through a converging lens and focused on the FF sample. The scattered light at angle θ = 15° passed through a diaphragm and was detected by a photomultiplier. The signal from the photomultiplier arrived to a computer for the correlation analysis.

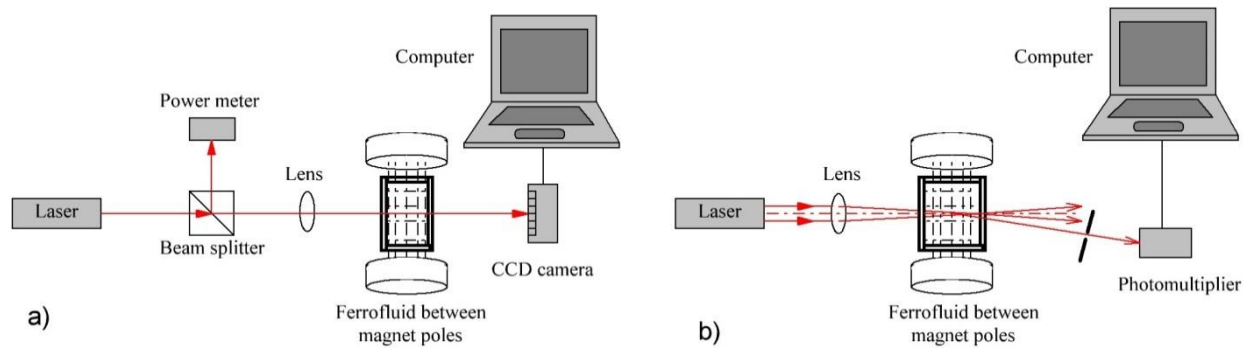

**Fig. 1.** Experimental setups for measuring scattered radiation from a ferrofluid sample: a – direct measurement of laser radiation scattering; b – laser correlation spectroscopic technique

Radiation sources were chosen so that the extinction in the samples was not too high. To this end, their optical transmission lines were measured in advance. A He-Ne laser with wavelength λ = 632.8 nm, a high stability and narrow spectral line was chosen for laser correlation measurements. In a number of experiments with direct scattering detection a more powerful Nd:YAG laser with the generation at the second harmonic, λ = 532 nm was used.

### 3. Results

#### 3.1. Direct measurements of scattered radiation

In the experiments involving direct measurements of scattered radiation, the effect of external magnetic field on the scattered light pattern was examined. At sufficiently high magnetic field, the pattern consisted of spots and was elongated in the direction perpendicular to the applied field. The locations of the spots had a certain periodicity. An example of this effect is illustrated in Fig. 2 which shows the pattern of radiation scattered by a kerosene-based sample with a concentration of magnetic nanoparticles of 0.2% at H = 1 kOe. The pattern was photographed by a CCD camera. Note that during the magnetic field rotation the scattering pattern followed it.

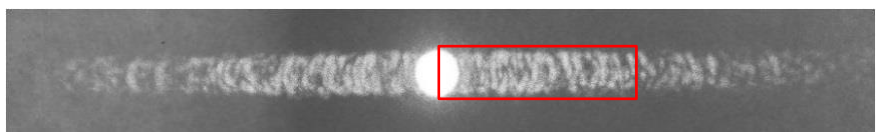

**Fig. 2.** Light scattering pattern for kerosene-based sample with a concentration of solid fraction of 0.2% registered for the incident light at λ = 532 nm and radiation power of 14 mW. The cell thickness is 0.1 mm, H = 1 kOe. The fragment which was subsequently analyzed by fast Fourier transformation is marked by a rectangle

Analysis of the images similar to those shown in Fig. 2 was carried out with the help of fast Fourier transformation (FFT) performed with specialized software *Image J*. The result of the processing of the part of light scattering pattern, marked by rectangle in Fig. 2 is shown in Fig. 3. Fig 3 (a) and Fig 3 (b) present the power spectra obtained by scanning the image along the field and perpendicular to it, respectively. Fig 3 (a) is a typical diffraction pattern with a sufficiently well resolved peaks. In contrast, as shown in Fig 3 (b), the resolution for the perpendicular direction is much worse. Thus the image represents the diffraction from certain anisotropic scatterers. Most likely, this is the manifestation of the emergence of the structure consisting of highly elongated regularly spaced agglomerates arising in the ferrofluid under the action of magnetic field [10].

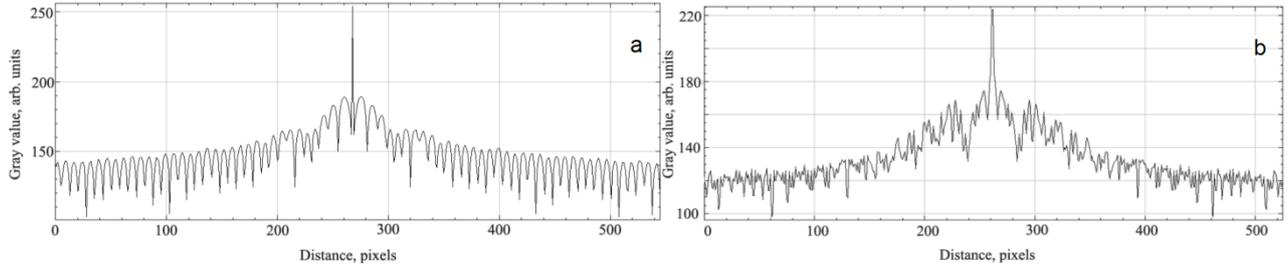

**Fig. 3.** FFT power spectra along the magnetic field (a) and in the perpendicular direction (b). The experimental conditions are the same as for Fig. 2, one pixel is approximately equal to 0.1 mm

The existence of these large clusters is well pronounced in high fields. It is more convenient to study their emergence and growth dynamics at low magnetic fields by the laser correlation method.

*3.2. Laser correlation spectroscopy*
The second method used to study particle agglomeration was Laser correlation spectroscopy. The autocorrelation function $g(\tau)$ was calculated from the scattering signals obtained at the setup shown in fig.1. $g(\tau)$ is typically characterized by an exponential behavior with the power depending on the particle size

$$|g(\tau)| = \int_0^\infty F(D)e^{-Dq^2\tau}dD, \quad (1)$$

where D is the diffusion coefficient, and F(D) is the contribution of the radiation component scattered by the particles of one size to the total intensity, $q = (4\pi n/\lambda)\sin(\theta/2)$ is the scattering vector, n is the refractive index of the medium, $\lambda$ is the wavelength, and $\theta$ is the angle of scattering. By using regularization methods, it is possible to calculate the D coefficient and then, by using the Stokes-Einstein relation

$$D = k_b T/6\pi\eta R, \quad (2)$$

to calculate the radius of the particles (or their agglomerates) under study. Here, $\eta$ is the viscosity of the medium, $k_b$ is the Boltzmann constant, $T$ is the temperature, and $R$ is the particle radius. The temperature was constant within the laser spot.
In our study the original algorithm for scatterers sizing was developed. The algorithm allows one to achieve the accuracy of determination of particle (or agglomerate) sizes in polydisperse solutions of up to 0.5 nm [11]. In order to achieve an acceptable transparency, the samples were diluted to a concentration of 0.018 vol. % of $Fe_3O_4$. As examples, the size distributions of magnetic particles obtained for zero magnetic field and magnetic field of 10 Oe are shown in Fig. 4.

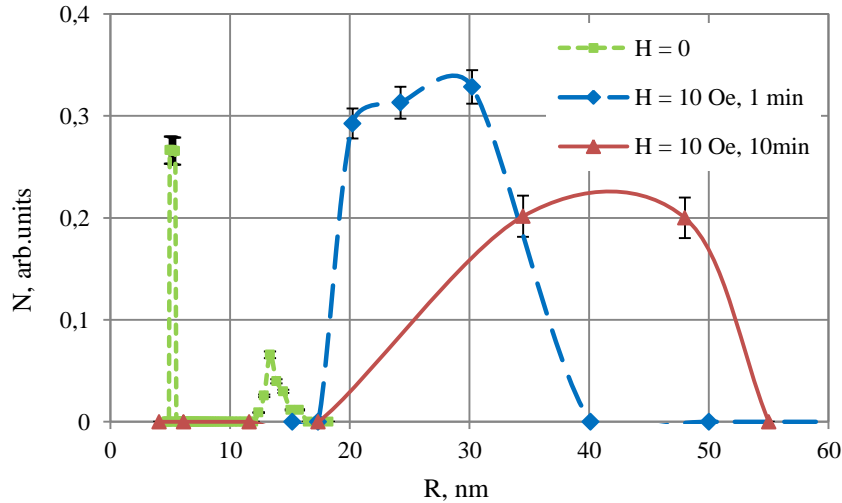

**Fig. 4.** Size distribution of magnetic particles (or/and their agglomerates) under the action of magnetic field (the experimental data are presented by the smoothed function)

The sizes of magnetic particles in the ferrofluid samples were also investigated by means of electron microscopy. It was found that the radii of individual particles did not exceed 7 nm. Note, however, that, as can be seen from Fig. 4, even in the absence of magnetic field a certain percentage of particles appear to be aggregated. This is reflected by the presence of an additional peak in the size distribution at about 13 nm which becomes more pronounced after the sample is exposed to a laser beam during several minutes. This suggests that the incident optical radiation interacts with the nanoparticles, which results in cluster formation. The mechanism of this interplay is not well understood now. According to [12], the photoinduced polarizability of magnetite can change in the presence of radiation, which can provoke an instability of a colloidal system. Another possibility is a partial aggregation of particles over the long storage time (it is an important issue to be considered for the industrial use of FFs).

Under the action of magnetic field the sizes of the scattering objects in the sample significantly increase, up to 20 – 35 nanometers, and grow to almost 50 nm after 10 minutes. Apparently in this case magnetic particles form conglomerates. Thus, it may be concluded that the laser correlation spectroscopy is an efficient tool for studying the structures of magnetic particles in FF under the influence of an external magnetic field and also in the presence of optical radiation.

**4. Conclusions**

The results of studies of magnetic properties of ferrofluids by two optical methods are discussed. It has been shown that the observation of the laser light scattered by the magnetic fluids allows one to estimate structure formed in a fluid by aggregates of nanoparticles. More detailed information related to the formation of clusters under the action of low external magnetic fields can be given by laser correlation spectroscopy. The latter allowed us to estimate the sizes of the objects formed under the action of magnetic field and also to observe and describe (preliminarily) the dynamics of their growth.

**Acknowledgments**

The authors are grateful to E.E. Bibik for providing ferrofluid samples and A.V. Varlamov for the help in experiments.